\documentclass[12pt]{article}

\newcommand{\Om}{\Omega}

\newcommand{\Tr}{\rm{Tr}}

\newcommand{\SG}{{S_{G, {\rm{symp}}}}}

\newcommand{\V}{{\cal V_{\rm symp}}}

\newcommand{\br}{{\bf R}}

\usepackage{latexsym,amsfonts,amsmath,amsthm,amssymb,graphicx}

\usepackage[T1]{fontenc}
\usepackage[latin1]{inputenc}
\title{Quantum mechanics from time scaling and random 
fluctuations at the ``quick time scale''}
\author{Andrei Khrennikov\footnote{Supported in part by the EU Human
Potential Programme, contact HPRN--CT--2002--00279 (Network on
Quantum Probability and Applications) and Profile Math. Modelling in
Physics and Cogn. Sc. of V\"axj\"o University.} \\International
Center for Mathematical
Modeling \\ in Physics and Cognitive Sciences,\\
University of V\"axj\"o, S-35195, Sweden\\
Email:Andrei.Khrennikov@msi.vxu.se}

\begin{document}
\date{}

\maketitle

\abstract{We show that (in contrast to a rather common opinion) QM
is not a complete theory. This is a statistical approximation of
classical statistical mechanics on the {\it infinite dimensional
phase space.} Such an approximation is based on the asymptotic
expansion of classical statistical averages with respect to a small
parameter $\kappa.$ Therefore statistical predictions of QM are only
approximative and a better precision of measurements would induce
deviations of experimental averages from quantum mechanical ones. In
this note we present a natural physical interpretation of $\kappa$
as the time scaling parameter (between quantum and prequantum
times).}

\section{Introduction}

In [1] we  showed that the conventional quantum formalism can be
obtained as a statistical approximation of classical statistical
mechanics on the {\it infinite dimensional phase space.} Such an
approximation is based on the asymptotic expansion of classical
statistical averages with respect to a small parameter $\kappa.$ By
representing points of the phase space ("hidden variables") by
classical vector fields, $\psi(x)=(q(x),p(x)),$ we can interpret our
prequantum theory as a field theory, {\it prequantum classical
statistical field theory - PCSFT.}

We emphasize the basis of PCSFT is given by  the model of
"nonempty vacuum" inducing so called {\it vacuum fluctuations,} cf. 
SED, and different variants of stochastic QM and random field theory
[2]--[11] (cf. also with "prespace" considered by B. Hiley [12] and in
the probabilistic framework in [13]). The author was also strongly
influenced by papers [14]-[18], and especially the recent review [19]. 

In the present paper we interpret the small parameter $\kappa$ as
the {\it time scaling parameter.} This gives the possibility to
couple our model with studies in general relativity,  string theory
and cosmology on the structure of space-time on the Planckian scale.

To choose the small parameter of our model, $\kappa,$ we should choose quantum and prequantum 
time scales. There are a few different possibilities and we shall use one of them.
We choose the {\it atom time}-scale in QM and the {\it Planck time}-scale in the prequantum 
classical theory. At the atom scale we consider the ``Hartree time'':
$$
t_{q}=t_{H} =\frac{\hbar}{E_H},
$$
where
$
E_H
$
is the Hartree energy. Under such a choice of time scales,  our small parameter is given by
\begin{equation}
\label{SKA} \kappa= \frac{t_P}{t_{H} }\approx
10^{-27}. \end{equation} 

The choice of the atom time-scale is motivated by the fact that this is the 
characteristic time scale of atom physics, physics in modern laboratories. We can call
this scale observational time scale. The choice of a prequantum 
time scale is an essentially more 
complicated problem. At the present time we are not able to observe 
directly prequantum fluctuations. Thus we do not know the characteristic scale of those fluctuations.
We can only speculate. The choice of the Planck time-scale is just one of such speculations.

We show that QM can be interpreted as an
approximative description of physical reality which is obtained
through neglecting by time intervals $\Delta t \leq t_P.$ We
assume that such time intervals are negligibly small for some
observer (so $\Delta t \approx 0).$ Then  contributions of such a
magnitude into averages can be neglected.  And, hence, there can be
used the conventional quantum mechanical rules for computing of
averages.

One of the main consequences of our approach for cosmology is that
(in contrast to the very common opinion) the Planck time $t_P$ is
not at all the {\it "ultimate limit to our laws of physics"} (in the
sense of laws of classical physics). A rather common idea that only
the quantum description is possible at the Planck scale for time and
space should be completed by the following thesis: "Yes, the quantum
description is possible, but it should be considered not as an
alternative to the classical one, but as an approximation."

The crucial point is that in our framework "classical" is not at all
"usual classical" (i.e., classical statistical mechanics on the
phase space $\Omega=\br^3 \times \br^3),$ but "infinite dimensional
classical". Thus at the Planck scale physical space is {\it
fundamentally infinite dimensional.} In our approach the
conventional mathematical model of physical space, namely $\br^3$,
should be rejected. We proceed in the similar way as it was done in
string theory, but, instead of finite dimensional models, e.g.,
$\br^{26},$ we should consider {\it infinite dimensional physical
space.}

We hope that our approach  may clarify the situation with
constructing quantum gravity. It is well known that, in spite of a
partial success, one could not say that the "project quantum
gravity" was completely successful. Our approach says that such a
rather negative result might be expected.  By our ideology "quantum"
$\sim$ "approximative". But, if general relativity is really a
complete (classical) physical theory, quantum general relativity has
no meaning. Even if general relativity is not complete theory and a
finer description is possible, it is more natural to assume that
such a pre general relativity would be a new classical theory.

Mathematically our model is based on {\it the Brownian motion in the space 
of classical fields.} The averaging with respect to a quantum state 
given by the von Neumann density operator $D$ can be represented as an approximation of 
averaging with respect to the classical random field given by the Wiener process 
(with the covariance operator $D)$ taking values in the space of classical fields.\footnote{The
ordinary Wiener process is a map $(t, \omega) \to w(t,\omega)\in {\bf R}^3,$  
where $t$ and $\omega$ are time and chance parameters, respectively. We consider the field-valued
Wiener process which is a map $(t, \omega) \to w(t,\omega)\in L_2({\bf R}^3).$ For each 
$(t, \omega),$ the classical field $\psi=w(t,\omega)$ belongs to the $L_2$-space:
$\psi= \psi(t,\omega,x)$ is a random field.} 
One can imagine quantum randomness 
as purely classical randomness which is similar to randomness of a Brownian particle.
The only difference is that, instead of motion of a particle colliding with molecules, we
consider motion of a classical field colliding with  random field media (fluctuations of
the background field). The crucial point is that such a Brownian motion of classical 
fields is performed on a {\it very quick time scale} which is essentially finer than time scales
approachable in modern physics. We are not able to observe this random motion directly 
(as we can do for the conventional Brownian motion of particles). Nevertheless, one could 
expect that sooner or later such a fine time scale would be approached and e.g. electron would be 
seen as a fluctuating classical field (interacting with the background field). Our model supports
expectations that it would be possible to go beyond QM.

We point out to the following terminological problem. Typically the Planck scale 
is regarded as quantum scale. There are numerous speculations that at this scale 
there is no more place for classical physics and the quantum description of phenomena 
is the only possible one. PCSFT does not support such views. From the point of view of PCSFT
QM is an approximative theory. Therefore it is natural to define quantum scale 
as the scale of applicability of QM. In our approach this is the scale of quantum 
systems: atoms, electrons, mesons, neutrino,... This scale is far from the Planck scale.
Thus the Planck scale has nothing to do with really quantum scale. Moreover, PCSFT does 
support the viewpoint that the Planck scale might be appropriative for description of
{\it prequantum fluctuations.} Therefore we choosen this scale as a {\it prequantum scale.}
Hence, in short:

\medskip

 {\it  In this paper the atom scale is called qauntum, the Planck scale is called 
prequantum.}

\section{On the properties of classical$\to$quantum correspondence}

We define {\it ``classical statistical models''} in the following
way, see [1] for more detail (and even philosophic considerations):
a) physical states $\omega$ are represented by points of some set
$\Omega$ (state space); b) physical variables are represented by
functions $f: \Omega \to {\bf R}$ belonging to some functional space
$V(\Omega);$ c) statistical states are represented by probability
measures on $\Omega$ belonging to some class $S(\Omega);$ d) the
average of a physical variable (which is represented by a function
$f \in V(\Omega))$ with respect to a statistical state (which is
represented by a probability measure  $\rho \in S(\Omega))$ is given
by
\begin{equation}
\label{AV0} < f >_\rho \equiv \int_\Omega f(\psi) d \rho(\psi) .
\end{equation}
A {\it classical statistical model} is a pair $M=(S, V).$\footnote{
We recall that classical statistical mechanics on the phase space
$\Omega_{2n}= {\bf R}^n\times {\bf R}^n$ gives an example of a
classical statistical model. But we shall not be interested in this
example in our further considerations. We shall develop  a classical
statistical model with {\it an infinite-dimensional phase-space.}}

The conventional quantum statistical model with the complex Hilbert
state space $\Omega_c$ is described in the following way: a)
physical observables are represented by operators $A: \Omega_c \to
\Omega_c$ belonging to the class of continuous self-adjoint
operators ${\cal L}_s \equiv {\cal L}_s (\Omega_c);$ b) statistical
states are represented by von Neumann density operators (the class
of such operators is denoted by ${\cal D} \equiv {\cal D}
(\Omega_c));$ d) the average of a physical observable (which is
represented by the operator $A \in {\cal L}_s (\Omega_c))$ with
respect to a statistical state (which is represented
  by the density operator $D \in {\cal D} (\Omega_c))$ is given by von Neumann's
formula:
\begin{equation}
\label{AV1} <A >_D \equiv \rm{Tr}\; DA
\end{equation}
The {\it quantum statistical model} is the pair $N_{\rm{quant}}
=({\cal D}, {\cal L}_s).$

\medskip

We are looking for a classical statistical model $M=(S, V)$ which
will provide {\it ``dequantization'' of the quantum model} $N_{\rm{quant}}
=({\cal D}, {\cal L}_s).$ By dequantization we understand constructing of 
a classical statistical model such that averages given by this model can be approximated
by quantum averages. Approximation is based on the asymptotic expansion of 
classical averages with respect to a small parameter. The main term of this expansion 
coincides with the corresponding quantum average.

One should not mix 
dequantization with so called {\it deformation quantization,} see e.g. [20]. The fundamental 
principle of deformation quantization is so called {\it correspondence principle.} By this 
principle if one considers the Planck constant $h$ as a small parameter then in the 
$h \to 0$ quantum mechanics is transformed into classical mechanics on the 
finite-dimensional phase space -- classical mechanics of particles. In this framework 
quantum mechanics is considered as a theory which is more general (``precise'') than 
classical mechanics of particles. Only by neglecting by actions of the Planck magnitude
one can apply the classical description. If actions of the Planck magnitude are 
taken into account then the classical mechanical description of phenomena is not more valid.
Classical predictions (for statistics of particles) can deviate from quantum predictions. The latter are considered 
as totally precise. Such deviations can be tested experimentally.

Dequantization should not be considered as the inverse operation to deformation quantization.
One point is that a classical  prequantum  model
(which we will construct) does not coincide with the conventional classical  statistical mechanics
on the finite dimensional phase space. Our prequantum model is not classical statistical mechanics 
of  particles, but it is statistical mechanics of classical  fields. Thus phase space has the infinite 
dimension. Therefore deformation quantization and dequantization simply operate on totally 
different phase spaces.

Another point (which is the crucial one) is that in the dequantization framework 
not quantum mechanics, but prequantum classical 
statistical model provides a better description of physical phenomena. 
The quantum description is only an approximative description. It can be applied only if one neglects by some small parameter
(which is time scale parameter $\kappa$ in this paper). By taking into account this parameter 
one obtains a better description of physical phenomena. Predictions 
of quantum mechanics should be violated. Those violations might be tested experimentally.

We remark that the procedure of dequantization does not contradict to the procedure of 
deformation quantization.\footnote{Although we do not want to critisize the correspondence 
principle, we should remark that this principle (which was elaborated by N. Bohr) has never 
beeen totally justified.}
 These are two different, but consistent limiting procedures.
We have the following scale of statistical descriptions of physical reality:
\begin{equation}
\label{DG}
\rm{Classical \;field\; theory} \to \rm{Quantum\;\;mechanics}
\end{equation}
\begin{equation}
\label{DG1}
\to
\rm{Mechanics\; \;of\; classical \; particles}
\end{equation}

In fact, all ``NO-GO'' theorems (e.g., von
Neumann, Kochen-Specker, Bell,...) can be interpreted as theorems
about impossibility of various dequantization procedures. Therefore
we should define the procedure of dequantization in such a way that
there will be no contradiction with known ``NO-GO'' theorems, but
our dequantization procedure still will be natural from the physical
viewpoint. We define (asymptotic) dequantization as a family
$M^\kappa=(S^\kappa, V)$ of classical statistical models depending
on small parameter $\kappa \geq 0.$ There  should exist maps
$T:S^\kappa\to {\cal D}$ and $T: V \to {\cal L}_s$ such that: a)
both maps are {\it surjections} (so all quantum states and
observables can be represented as images of classical statistical
states and variables, respectively); b) the map $T: V \to {\cal
L}_s$ is ${\bf R}$-linear (we recall that we consider real-valued
classical physical variables); c) the map $T:S\to {\cal D}$ is
injection (there is one-to one correspondence between classical and
quantum statistical states); d) classical and quantum averages are
coupled through the following asymptotic equality:
 \begin{equation}
\label{AQ} < f >_\rho = \kappa <T(f)>_{T(\rho)} + O(\kappa^2), \; \;
\kappa \to 0
\end{equation}
(here $<T(f)>_{T(\rho)}$ is the  quantum average); so:
\begin{equation}
\label{AQ1}
\int_\Omega f(\psi) d \rho(\psi)=  \kappa \; \Tr \; D A
+ O(\kappa^2), \; \; A=T(f), D= T(\rho).
\end{equation}
This equality can be interpreted in the following way. Let $f(\psi)$
be a classical physical variable (describing properties of
microsystems - classical fields having very small magnitude
$\kappa).$  We define its {\it amplification} by:
 \begin{equation}
\label{AMP} f_\kappa(\psi) =\frac{1}{\kappa} f(\psi)
\end{equation}
(so any micro effect is amplified in $\frac{1}{\kappa}$-times).
Since $\kappa$ gives {\it intensity of vacuum fluctuations,} the
quantity $f_\kappa$ can be interpreted as relative intensity of $f$
with respect to vacuum fluctuations. For such a relative intensity,
we have:
\begin{equation} \label{AQ4}
< f_\kappa >_\rho = <T(f)>_{T(\rho)} + O(\kappa), \; \; \kappa \to
0.
\end{equation}

Hence: {\it QM is a mathematical formalism describing a statistical
approximation of relative intensities of classical field variables
with respect to vacuum fluctuations.} \footnote{We recall that in
the von Neumann ``NO-GO'' theorem there was assumed that the
correspondence $T$ between classical variables and quantum
observables is {\it one-to-one.} Thus our dequantization violates
this von Neumann condition. Therefore the von Neumann theorem could
not be applied to PCSFT. On the other hand, the map $T: V\to {\cal
L}_s$ (given by (\ref{Q30})) is ${\bf R}$-linear as it was
postulated by J. von Neumann  (thus, e.g., $T(f_1+f_2)=
T(f_1)+T(f_2)$ even in the case of noncommuting operators $T(f_1)$
and $T(f_2)).$ We recall that this assumption was criticized by many
authors, in particular, by J. Bell. The crucial difference with
dequantizations considered in known ``NO-GO'' theorems is that in
our case {\it classical and quantum averages are equal only
asymptotically.} We also remark that a classical variable $f$ and
the corresponding quantum observable $A=T(f)$ can have {\it
different ranges of values.} In particular, the latter possibility
blocks application of Bell's theorem to PCSFT.}

\section{Prequantum classical statistical field theory}

We choose the phase space $\Om= Q\times P,$ where $Q=P=H$ and $H$ is
the infinite-dimensional real (separable) Hilbert space. We consider
$\Omega$ as the real Hilbert space with the scalar product $(\psi_1,
\psi_2)= (q_1, q_2) + (p_1, p_2).$ We denote  by $J$ the symplectic
operator on $\Omega:
 J= \left( \begin{array}{ll}
 0&1\\
 -1&0
 \end{array}
 \right ).$
Let us consider the class ${\cal L}_{\rm symp} (\Omega)$ of bounded
${\bf R}$-linear operators $A: \Omega \to \Omega$ which commute with
the symplectic operator:
A J= J A .
This is a subalgebra of the algebra of bounded linear operators
${\cal L} (\Omega).$ We also consider the space of ${\cal
L}_{\rm{symp}, s}(\Omega)$ consisting of self-adjoint operators.

By using the operator $J$ we can introduce on the phase space
$\Omega$ the complex structure. Here $J$ is realized as $-i.$ We
denote $\Omega$ endowed with this complex structure by $\Omega_c:
\Omega_c\equiv Q\oplus i P.$ We shall use it later. At the moment
consider $\Omega$ as a real linear space and consider its
complexification $\Omega^{{\bf C}}= \Omega \oplus i \Omega.$

Let us consider the functional space ${\cal V}_{\rm{symp}}(\Omega)$
consisting of functions $f:\Omega \to {\bf R}$ such that: a) the
state of vacuum is  preserved : $f(0)=0;$ b) $f$ is $J$-invariant:
$f(J\psi)= f(\psi);$ c) $f$ can be extended to the  analytic
function $f:\Omega^{{\bf C}}\to {\bf C}$ having  the exponential
growth: $ \vert f(\psi)\vert \leq c_f e^{r_f \Vert \psi \Vert}
$
for some $c_f, r_f \geq 0$ and for all $\psi\in \Omega^{{\bf C}}.$
\footnote{We remark that the possibility to extend a function $f$
analytically onto $\Omega^{{\bf C}}$ and the exponential estimate on
$\Omega^{{\bf C}}$ plays the important role in the asymptotic
expansion of Gaussian integrals. To get a mathematically rigor
formulation, conditions in [1] should be reformulated in the similar
way.}

The following trivial mathematical result plays the fundamental role
in establishing classical $\to$ quantum correspondence: {\it Let $f$
be a smooth $J$-invariant function. Then } $f^{\prime \prime}(0)\in
{\cal L}_{\rm{symp}, s}(\Omega).$ In particular, a quadratic form is
$J$-invariant iff it is determined by an operator belonging to
${\cal L}_{\rm{symp}, s}(\Omega).$

We consider the space statistical states $S_{G,
\rm{symp}}^{\kappa}(\Omega)$ consisting of measures $\rho$ on
$\Omega$ such that: a) $\rho$ has zero mean value; b) it is a
Gaussian measure; c) it is $J$-invariant; d) its dispersion has the
magnitude $\kappa.$ Thus these are $J$-invariant Gaussian measures
such that $$ \int_\Omega \psi d\rho(\psi)=0 \; \mbox{and}\;
\sigma^2(\rho)= \int_\Omega \Vert \psi\Vert^2 d \rho(\psi)= \kappa,
\; \kappa \to 0.
$$
Such measures describe small Gaussian fluctuations of the vacuum
field. The following trivial mathematical result plays the
fundamental role in establishing classical $\to$ quantum
correspondence: {\it Let a measure $\rho$ be $J$-invariant. Then its
covariation operator} $B= \rm{cov}\; \rho \in {\cal L}_{\rm{symp},
s}(\Omega).$ Here $(By_1, y_2)= \int (y_1, \psi)(y_2, \psi) d \rho(
\psi).$

We now consider the complex realization $\Omega_c$ of the phase
space and the corresponding complex scalar product $<\cdot, \cdot>.$
We remark that the class of operators ${\cal L}_{\rm symp} (\Omega)$
is mapped onto the class of ${\bf C}$-linear operators ${\cal
L}(\Omega_c).$ We also remark that, for any $A\in {\cal
L}_{\rm{symp}, s}(\Omega),$ real and complex quadratic forms
coincide:
$ (A\psi,\psi) =<A\psi,\psi>.$
We also define for any measure its complex covariation operator
$B^c= \rm{cov}^c \rho$ by $ <B^c y_1, y_2>=\int <y_1, \psi> <\psi,
y_2> d \rho (\psi). $ We remark that for a $J$-invariant measure
$\rho$ its complex and real covariation operators are related as
$B^c=2 B.$ As a consequence, we obtain that any $J$-invariant
Gaussian measure is uniquely determined by its complex covariation
operator. As in the real case [1], we  can prove that for any
operator $ A\in {\cal L}_{\rm{symp}, s}(\Omega):$
$\int_\Omega <A\psi,\psi> d \rho (\psi) = \rm{Tr} \;\rm{cov}^c \rho
\;A.$
 We pay attention that the trace is considered with respect to the complex
inner product.

 We consider now the one parameter family of classical statistical
models:
\begin{equation}
\label{MH} M^\kappa= ( S_{G, \rm{symp}}^\kappa(\Omega),{\cal
V}_{\rm{symp}}(\Omega)), \; \kappa\geq 0,
\end{equation}

By making in the Gaussian infinite-dimensional integral the change
of variables (field scaling):
\begin{equation}
\label{ANN3Z} \psi= \sqrt{\kappa} \Psi,
\end{equation}
we obtain the following result:

\medskip

{\bf Lemma 1.} {\it Let $f \in {\cal V}_{\rm{symp}}(\Omega)$ and let
$\rho \in S_{G, \rm{symp}}^\kappa(\Omega).$ Then the following
asymptotic equality holds:
\begin{equation}
\label{ANN3} <f>_\rho =  \frac{\kappa}{2} \; \rm{Tr}\; D^c \;
f^{\prime \prime}(0) + O(\kappa^2), \; \kappa \to 0,
\end{equation}
where the operator $D^c= \rm{cov}^c \; \rho/\kappa.$ Here
\begin{equation}
\label{OL} O(\kappa^2) = \kappa^2 R(\kappa, f, \rho),
\end{equation}
where $\vert R(\kappa,f,\rho)\vert \leq c_f\int_\Omega  e^{r_f \Vert
\Psi \Vert}d\rho_{D^c} (\Psi).$ }

\medskip
Here $\rho_{D^c}$ is the Gaussian measure with zero mean value and
the complex covariation operator $D^c.$

We see that the classical average (computed in the model $M^\kappa=
( S_{G, \rm{symp}}^\kappa(\Omega),{\cal V}_{\rm{symp}}(\Omega))$ by
using the measure-theoretic approach) is coupled through
(\ref{ANN3}) to the quantum average (computed in the model
$N_{\rm{quant}} =({\cal D}(\Omega_c),$ ${\cal L}_{{\rm
s}}(\Omega_c))$ by the von Neumann trace-formula).

The equality (\ref{ANN3}) can be used as the motivation for defining
the following classical $\to$ quantum map $T$ from the classical
statistical model $M^\kappa= ( S_{G, \rm{symp}}^\kappa,{\cal
V}_{\rm{symp}})$ onto the quantum statistical model
$N_{\rm{quant}}=({\cal D}, {\cal L}_{{\rm s}}):$
\begin{equation}
\label{Q20} T: S_{G, \rm{symp}}^\kappa(\Omega) \to {\cal
D}(\Omega_c), \; \; D^c=T(\rho)=\frac{\rm{cov}^c \; \rho}{\kappa}
\end{equation}
(the Gaussian measure $\rho$ is represented by the density matrix
$D^c$ which is equal to the complex covariation operator of this
measure normalized by  $\kappa$);
\begin{equation}
\label{Q30} T: {\cal V}_{\rm{symp}}(\Omega) \to {\cal L}_{{\rm
s}}(\Omega_c), \; \; A_{\rm quant}= T(f)= \frac{1}{2}
f^{\prime\prime}(0).
\end{equation}
Our previous considerations can be presented as

\medskip

{\bf Theorem 1.} {\it The one parametric family of classical
statistical models $M^\kappa= ( S_{G,
\rm{symp}}^\kappa(\Omega),{\cal V}_{\rm{symp}}(\Omega))$ provides
dequantization of the quantum model $N_{\rm{quant}} =({\cal
D}(\Omega_c),$ ${\cal L}_{{\rm s}}(\Omega_c))$ through the pair of
maps (\ref{Q20}) and (\ref{Q30}). The classical and quantum averages
are coupled by the asymptotic equality (\ref{ANN3}).}

\section{Time-scaling for infinite-dimensional Wiener process}

Let $w_s^D, s \geq 0,$ be the $\Omega$-valued Wiener process
corresponding to the trace class (self-adjoint) operator $D \geq 0$
with $\Tr D=1.$  We also assume that $D$ is $J$-commuting. Thus
\begin{equation}
\label{WP1}
E < \phi, w_s^D>=0, \phi \in \Omega,
\end{equation}
\begin{equation}
\label{WP2}
E < \phi_1, w_s^D><w_s^D, \phi_2>= s <D\phi_1, \phi_2>, \phi_1, \phi_2 \in \Omega.
\end{equation}
Then we have:
\begin{equation}
\label{WP}
{\rm Prob.\; law}\; (w_{\kappa s}^D: s \geq 0) =\\
{\rm Prob. \;law} \;(\kappa^{1/2} w_s^D: s \geq 0)
\end{equation}
for any $\kappa > 0.$  We shall see that by (\ref{WP}) our
$\kappa^{1/2}$-scaling of $\psi \in \Omega$ can be considered as the
result of $\kappa$-scaling of time.

\medskip

Our basic postulate is that {\it quantum formalism arises as the
result of an approximation based on the time scaling}

\medskip
Let us consider a {\it "prequantum time  scale"} that is essentially
finer\footnote{The meaning of "essentially" would be discussed
later.} than the quantum time scale. We suppose that these two time
scales can be coupled through a small scaling parameter $\kappa >
0.$ Denote the prequantum and quantum times by symbols $s$ and $t$
respectively. We suppose that:
\begin{equation}
\label{PT}
t= \kappa s
\end{equation}
Here $\kappa$ is a dimensionless parameter. It is assumed that
\begin{equation}
\label{TP} \kappa < < 1
\end{equation}
Thus the unit interval $s=1$ of the pre-quantum time corresponds to
a small interval $t=\kappa$ of the quantum time. We can also say
that the unit interval $t=1$ of quantum time corresponds to a huge
interval $s=\frac{1}{\kappa}$ of the prequantum time. Moreover, if
$\kappa \to 0,$ then $s=\frac{1}{\kappa} \to \infty.$ Therefore at
the prequantum time scale quantum processes have practically
infinite duration.

Let us consider the time scaling (\ref{PT}) for the Wiener processes
$w_s^D.$  We set
$$W_t^D=w_{\kappa s}^D.$$
The formula (\ref{WP}) implies that, for any continuous function $f:
\Omega \to \br$ (which is integrable with respect to any Gaussian
measure on $\Omega$), we have:
\begin{equation}
\label{WP1Z} E f(W_\kappa^D)= E f(\kappa^{1/2} w_1^D).
\end{equation}
This is nothing else than the basic "field-scaling" formula
(\ref{ANN3Z})! We interpret $W_t^D$ as the Wiener process with
respect to the quantum time $t$ and $w_s^D$ as the Wiener process
with respect to the prequantum time $s$.

The tricky thing with the quantum formalism is that it does not give
a possibility to find exactly the average $E f(W_t^D)$ with respect
to the "quantum Wiener process" $W_t^D.$ The main problem is that
the interval $t=\kappa$ is negligibly small (from the QM-viewpoint).

The quantum formalism provides only an approximation of the
classical average $E f(W_\kappa^D).$

Moreover, to produce observable effects, the classical physical
variable $f$ should be amplified: $f \to f_\kappa \equiv
\frac{1}{\kappa} f.$

In such an approach to QM,  we can proceed through expanding the
right-hand side of (\ref{WP1}) into series with respect to the
scaling parameter $\kappa^{1/2}.$ If we take $f \in \V(\Omega)$ then
by using the probabilistic notations we can repeat considerations of
section 3:
\begin{equation}
\label{PE} E f_\kappa (W_\kappa^D)= E f_\kappa (\kappa^{1/2} w_1^D)=
\frac{1}{2} E (f^{\prime\prime} (0) w_1^D, w_1^D) + O(\kappa),
\kappa \to 0.
\end{equation}
Therefore, for nonquadratic maps $f: \Omega \to \br,$ QM gives only
an approximation $<f_\kappa>_D= \Tr Df^{\prime\prime}(0)$ of the
real average $E f_\kappa (W_\kappa^D).$

The difference between statistical predictions of QM and PCSFT is of
the magnitude $\kappa,$  where $\kappa$ is the scaling parameter
coupling the prequantum and quantum time scales, see (\ref{PT}).
What is a magnitude of the time scaling factor $\kappa$?

Thus by taking into account Brownian fluctuations on the prequantum
time scale we can say that prequantum statistical states are given
by Wiener measures $P_D$ on the space $C_0([0, \kappa]),$ of
trajectories $\psi: [0, \kappa] \to \Omega, \psi(0)=0.$ Denote this
space of such Wiener measures by the symbol $\SG(C_0([0, \kappa]),
\Omega)$ (we recall that $[D, J]=0).$

This is the space of statistical states of our new prequantum
classical statistical model. As the space of classical physical
variables, we should choose some subspace of the space of continuous
functionals $f: C_0([0, \kappa], \Omega) \to \br.$

Since all our considerations are coupled to the fixed moment of
(quantum) time $t=\kappa,$ we can restrict our considerations to the
class of functionals which depend only on $\psi(\kappa).$ So we can
choose the space of classical physical variables consisting of
functionals of trajectories, $\psi: [0, \kappa] \to \Omega,$ of the
form $\psi(\cdot) \to f(\psi(\kappa)), f \in \V(\Omega).$ We denote
this class by the symbol $\V(C_0([0, \kappa], \Omega)).$

Thus, finally, we consider the following classical statistical model
on phase space consisting of trajectories $\tilde \Omega^\kappa=C_0
([0, \kappa], \Omega):$
$$\tilde M^\kappa= (\SG(\tilde \Omega^\kappa), \V(\tilde \Omega^\kappa)).$$
We define the maps $T$ in the same way as in section 3:
\begin{equation}
\label{F1} T:\SG(\tilde\Omega^\kappa) \to {\cal D}(\Omega_c), \; \;
T(P_D)=D;
\end{equation}
\begin{equation}
\label{F2} T:\V (\tilde\Omega^\kappa)\to {\cal L}_s (\Omega_c),
 \; \; T(f)= f^{\prime\prime} (0)/2.
\end{equation}

{\bf Theorem 2.} {\it The family of  classical statistical model
$\tilde M^\kappa, \kappa > 0,$ and the pair of maps (\ref{F1}),
(\ref{F2}) provide dequantization of the conventional Dirac-von
Neumann quantum model $N_{\rm quant}$.}

\section{The magnitude of time scaling}

To get the small parameter of our model $\kappa,$ we should choose quantum and prequantum 
time scales. There are a few different possibilities and we shall discuss one of them.
We choose the {\it atom time}-scale in QM and the {\it Planck time}-scale in the prequantum 
classical theory. At the atom scale we consider the ``Hartree time'':
\begin{equation}
\label{T1}
t_{q}=t_{H}= \frac{\hbar}{E_H}
\end{equation}
Here
$$
E_H= \frac{\hbar^2}{m_e a_0^2}= m_e c^2 \alpha^2\approx 4.35974417(75)10^{-18} J 
$$
is the Hartree energy, $m_e$ is the electron mass, $a_0$ is the Bohr radius, and $\alpha$ is 
the fine structure constant.
Hence:
$$
t_{q}= \frac{\hbar}{m_e c^2 \alpha^2}\approx 2.418 8884 326505(16) \times  10^{-17} s
$$
And we have:
\begin{equation}
\label{T2}
t_{\rm{prq}}= t_P=\sqrt{\frac{\hbar G}{c^5}} \approx 5.39121(40) \times 10^{-44} s
\end{equation}
Therefore our time scaling parameter
\begin{equation}
\label{T3}
\kappa =\frac{t_{\rm{prq}}}{t_{q}}= \frac{t_P}{t_{H}}= \alpha^2 m_e \sqrt{\frac{G}{c\hbar}}=
\alpha^2 \frac{m_e}{m_P},
\end{equation}
where
$$
m_P=\sqrt{\frac{\hbar c}{G}}\approx 2.176\times 10^{-11} gr
$$
is the Planck mass. 
Thus our time-scaling parameter has the magnitude:
$$
\kappa \sim 10^{-27}.
$$

Under such a choice of the prequantum scale 
the difference  between statistical predictions of PCSFT and QM
(given by (\ref{PE}) is of the order $10^{-27}$. Thus, if, e.g., the
classical physical variable
$$f(\psi)= \frac{1}{2} <A \psi, \psi> +
\frac{1}{4} <A \psi, \psi>^2, A \in {\cal L}_s,$$ then the
difference between the quantum prediction ($<A>_D=\Tr DA$) and the
PCSFT-prediction should be of the order $10^{-27}.$ (Of course, under
the assumption that the Planck time $t_P$ really provides the
correct prequantum time-scale!)

\section{Links to general relativity, string theory, cosmology}

\subsection{Sub-Planckian time}
We see that in the PCSFT-framework it is possible to do classical
physics on sub-Planckian time scale\footnote{The main distinguishing
feature of such classical models is the infinite-dimension of phase
space.}. We recall that according to a rather common opinion the
Planck time is "the ultimate limit to our laws of physics". By this
point of view any classical description of physical reality is
impossible for the sub-Planckian time. Moreover, there is the
opinion that even the notion of time becomes meaningless for $s <
t_P,$ see, e.g., WikipediA (The free Encyclopedia) for this very
popular viewpoint: {\small "However, the Planck time $t_P$ may not
be taken as a "quantum of time." Within the framework of the laws of
physics as we understand them today, we can neither measure nor
discern any difference between the universe at the time it first
came into existence and the universe anything less than 1 Planck
time later."}

\medskip

In the opposition to this opinion, in our approach stochastic
dynamics on the sub-Planckian scale (dynamics with respect to the
"prequantum time" $s$) plays the fundamental role in reproduction of
quantum averages.

\subsection{Sub-Planckian space}
By considering the sub-Planckian  time scale we should also consider
the sub-Planckian length scale. However, this is not the case in our
model.\footnote{The conventional interpretation is that the Planck
time is the time it would take a photon traveling at the speed of
light to cross a distance equal to the Planck length. Our approach,
PCSFT, is a purely field model of reality, cf. Einstein [23].
Moreover, we consider fields not as "real waves" on the finite
dimensional physical space, but as points of the infinite
dimensional space. In PCSFT time is the evolutional parameter not
for dynamics in the conventional physical space, but in the infinite
dimensional Hilbert space. }

 In fact, by considering phase space $\Omega=Q \times P,$
where $Q=P=H$ is the real (separable) Hilbert space, we exclude from
the very beginning the physical space $\br^3$ from consideration. Of
course, one could choose the representation of $H$ by $L_2(\br^3),$
but we did this sometimes only to couple our model to physics of
classical fields. In principle, there are no reasons to introduce
the physical space into consideration.

This is a good place to discuss the role of physical space
represented by ${\bf R}^3$ in our model. In PCSFT the {\it real
physical space} is Hilbert space. If we choose the realization
$$
H=L_2({\bf R}^3),
$$
then we obtain the realization of $H$ as the space of classical
fields on ${\bf R}^3.$  So the {\it conventional space ${\bf R}^3$
appears only through this special representation of the Hilbert
configuration space.} Dynamics in ${\bf R}^3$ in just a shadow of
dynamics in the space of fields. However, we can choose other
representations of Hilbert configuration space. In this way we shall
obtain classical fields defined on other ``physical spaces.''

We remark that at first sight the situation with development of
PCSFT is somewhat reminiscent of the one confronted by Scr\"odinger
in his introduction of his wave equation, which ``maps'' waves in
the configuration space (the idea that in part derives from
Hamilton's mechanical-optical analogy that led Hamilton to his
version of classical mechanics). However, as is well known, in the
specific case considered by Scr\"odinger, the configuration space
and the physical space were both ${\bf R}^3,$ which coincidence was
in part responsible for Scr\"odinger's hope that his equation
describes an actual physical (wave) process in space-time. This hope
did not materialize, given that in general the configuration space
of physical systems is not
 ${\bf R}^3.$ The difficulties of Schr\"odinger's program quickly led to Born's
interpretation of the wave function in terms of probability or, as
Scr\"odinger himself came to call it ``expectation catalogue'',
which view is central in the orthodox or Copenhagen interpretation
of quantum mechanics. Even though he had, just as did Einstein,
major reservations concerning quantum mechanics as the ultimate
theory of quantum phenomena, Scr\"odinger never went so far as to
see any space other than  ${\bf R}^3$ as real.

The same can be said about Einstein's attempts to go beyond quantum
mechanics. His attempts, see, e.g., [23], to create purely field
model of physical reality did not induce rejection of the
conventional model of physical space. Nevertheless, some of his
comments might be interpreted as signs as coming rejection of the
conventional model of physical space, see [23]: {\small ``Space-time
does not claim existence on its own, but only as a structural
quality of the field.'' ``The requirement of general covariance
takes away from space and time the last remnant of physical
objectivity.''} And the following Einstein's remark is especially
important for PCSFT's view to physical space: {\small ``There is no
such thing as an empty space, i.e., a space without field.
Space-time does not claim existence on its own, but only as a
structural quality of the field.''} L. De Broglie in his theory of
double solution (the first hidden variable model) emphasized the
fundamental role of physical space ${\bf R}^3.$ 
Such a viewpoint also was common for adherents of Bohmian mechanics
(in any case for D. Bohm and J. Bell). We can conclude that all
former models with field-like hidden variables were based on the
conventional model of physical space, namely  ${\bf R}^3.$

On the other hand, string theory does introduce spaces of higher
dimensions, although not of infinite dimensions. This approach was
one of inspirations for our radical viewpoint to physical space. One
could speculate that on scales of quantum gravity and string theory
space became infinite dimensional, just as those theories the space
has the (finite) dimension higher than three. (In our approach
quantum theory is not the ultimate theory. It has its boundaries of
applications. Therefore there are no reasons to expect that
``quantum gravity'' should exist at all. Thus it would be better to
speak not about scales of quantum gravity, but simply about the
Planck scale for length and time).

Starting with classical statistical mechanics on the infinite
dimensional physical space (PCSFT), we first obtain quantum
mechanics and then classical statistical mechanics on the
finite-dimensional phase space:
\begin{equation}
\label{LLL} \lim_{h \to 0} \lim_{\kappa\to 0} \tilde M^\kappa=
\lim_{h \to 0}  N_{\rm{quant}}^h== M_{\rm{conv. class.}}.
\end{equation}

\section{Nonvacuum fluctuations}

The asymptotic equality (\ref{PE}) can be written  in the following
way (since we consider the class of functions with $f(0)=0):$
\begin{equation}
\label{E} \lim_{\kappa \to 0} E
\Big[\frac{f(W_\kappa^D)-f(0)}{\kappa}\Big] = <T(f)>_{T(P_D)}\equiv
\frac{1}{2} \Tr Df^{\prime\prime}(0) .
\end{equation}
Denote the average (classical!) $E f(\psi + W_\kappa^D)$ by the
symbol $\bar{f}(\kappa, \psi), \psi \in \Omega.$ We have:

\medskip

{\bf Proposition 1.} {\it The quantum average can be represented as
the derivative (at the vacuum point $\psi=0$) of the classical
average $\bar{f}(\kappa, \psi)$ at $\kappa=0:$}
\begin{equation}
\label{E1} \frac{\partial \bar{f}}{\partial \kappa}(0,
0)=<T(f)>_{T(P_D)}.
\end{equation}

\medskip

We pay attention that
\begin{equation}
\label{E2} {\cal U}(g)(\psi)=\frac{1}{2} \Tr Dg^{\prime\prime}(\psi)
\end{equation}
is the infinitesimal generator of the Wiener process $W_t^D,$ and
the function $\bar{f}(t, \psi)=Ef(\psi + W_t^D)$ is the solution of
the Cauchy problem:
\begin{equation}
\label{E3} \frac{\partial \bar{f}}{\partial t} (t, \psi)=\frac{1}{2}
\Tr D\; \bar{f}^{\prime\prime}(t, \psi),
\end{equation}
\begin{equation}
\label{E4} \bar{f}(0, \psi)=f(\psi)
\end{equation}
In the same way as we considered fluctuations of the vacuum field:
$\psi_{\rm vacuum} + W_t^D(\omega)\equiv W_t^D(\omega),$ we can
consider fluctuations of any field $\psi_0 \in \Omega: \psi_0 +
W_t^D (\omega).$ From the purely mathematical viewpoint, there is no
difference between such models. Since the differential operator
${\cal U}$ given by (\ref{E2}) is the infinitesimal generator, we
have the analogue of the asymptotic expansion for any $\psi_0 \in
\Omega:$
\begin{equation}
\label{E8}
E f_\kappa (\psi_0 + W_\kappa^D)= \frac{1}{2} E (f^{\prime\prime} (\psi_0) w_1^D, w_1^D) + O(\kappa), \kappa \to 0,
\end{equation}
for any $f$ such that $f_{\psi_0}(\psi)\equiv f(\psi-\psi_0) \in
\V(\Omega).$ Denote by $P_{\psi_0, D}$ the Wiener measure on the
space of trajectories $C_{\psi_0}([0, \kappa], \Omega)$
corresponding to the process $\psi_0 + W_t^D.$ Here the space
$C_{\psi_0}([0, \kappa], \Omega)$ consists of all continuous
trajectories $\psi: [0, \kappa] \to \Omega, \psi (0)=\psi_0.$ Denote
the space of such Wiener measures by the symbol $\SG(\tilde
\Omega_{\psi_0}^\kappa),$ where $\tilde \Omega_{\psi_0}^\kappa=
C_{\psi_0} ([0, \kappa], \Omega).$

We consider the classical statistical model corresponding to the
choice of $\psi_0$ as a (deterministic) background field: $$\tilde
M_{\psi_0}^\kappa= (\SG(\tilde\Omega_{\psi_0}^\kappa), \V(\tilde
\Omega_{\psi_0}^\kappa)).$$ Since, for any $f \in \V(\tilde
\Omega_{\psi_0}^\kappa),$ $$<f>_{P_{\psi_0,
D}}=\int_{\tilde\Omega_{\psi_0}^\kappa} f(\psi) dP_{\psi_0,
D}(\psi)=E f(\psi_0 + W_\kappa^D),$$ by (\ref{E8}) we have
\begin{equation}
\label{Z8}
<f_\kappa>_{P_{\psi_0, D}}=\frac{1}{2} \Tr Df^{\prime\prime}(\psi_0) + O(\kappa), \kappa \to 0.
\end{equation}
We modify the $T$-maps for classical statistical states and
variables:
\begin{equation}
\label{CSQ} T: \SG(\tilde \Omega_{\psi_0}^\kappa) \to {\cal
D}(\Omega_c), \; \; T(P_{\psi_0, D})=D;
\end{equation}
\begin{equation}
\label{CSQ1} T: \V(\tilde \Omega_{\psi_0}^\kappa) \to {\cal
L}_s(\Omega_c), \; \;   T(f)=f^{\prime\prime}(\psi_0)/2 .
\end{equation}

{\bf Theorem 3.} {\it For any $\psi_0 \in \Omega,$  the classical
statistical model $\tilde M^\kappa_{\psi_0}$ and the pair of maps
(\ref{CSQ}), (\ref{CSQ1}) provide dequantization of the conventional
Dirac-von Neumann quantum model $N_{\rm quant}$.}

\medskip

If we consider a function $f:\Omega \to \br$ such  that $f(\psi_0)
\ne 0,$ then (\ref{Z8}) is modified into
\begin{equation}
\label{Z9} \frac{\bar{f}(\kappa, \psi_0)-
f(\psi_0)}{\kappa}=\frac{1}{2} \Tr D\;
\bar{f}^{\prime\prime}(\psi_0) + O(\kappa), \kappa \to 0,
\end{equation}
or
\begin{equation}
\label{Z91} \frac{\partial \bar{f}(0, \psi_0)}{\partial \kappa}=
{\cal U}(\bar{f})(\psi_0) \equiv \frac{1}{2} \Tr D\;
\bar{f}^{\prime\prime} (\psi_0).
\end{equation}
Thus we found that the operator of quantum averaging  $f \to
\frac{1}{2} \Tr Df^{\prime\prime} (\psi_0)$ is nothing else than the
infinitesimal generator of the prequantum Wiener process.

\medskip

{\bf Remark.} (Why the vacuum background?) As we have seen  the same
quantum model can be derived starting with fluctuations of any
background field $\psi_0.$ The quantum model applied in physics only
for $\psi_0=0,$ because effects of small magnitudes could not be
extracted in the presence of a nontrivial background field $\psi_0.$

\section{Appendix: no-go theorems, PCSFT and nonlocality}

\subsection{Comparing with no-go theorems of von Neumann,
Cohen-Specker and Bell}

The referee of this paper made the following remark on PCST: \small{``The author 
attempts to derive quantum mechanics from a classical phase
space which is infinite dimensional.  If this is a correct theory it seems to
imply that quantum mechanics should be derivable from classical physics in the
limit of an infinite number of particles. A classical system of $N$
non-relativistic particles moving in
3-dimensional Euclidean space has a phase space which is ${\bf R}^{6N}$.  As
$N\rightarrow\infty$ this would seem to agree with the author's starting
point.  If this is the case then
quantum mechanics would be derivable from classical fluid dynamics and I do
not believe that this is possible.''} The crucial problem in such a discussion is 
interpreting the word {\it ``derive.''} In fact, we do not claim that QM can be 
reproduced as a part of classical phase space mechanics. We study possibilities to {\it map}
the classical phase space mechanics onto QM. As we have seen, our classical$\to$ quantum mapping 
is a rather special -- in particular, it is not one-to-one,-  but, nevertheless, there is 
an asymptotic correspondence bewteen classical and quantum averages.

However, there are no-go theorems for mathematical attempts to have a map
from classical variables to quantum operators which preserves
statistics, e.g., theorems of von Neumann, Cohen-Specker and Bell. 
The no-go theorems say: No such map exists. In this paper
we constructed such a map. What goes?

Our construction does not contradict to known no-go theorems, since
our map $T$ does not satisfy some conditions of those theorems. An
important condition in all such theorems is that the {\it range of
values} of a classical variable $f$ should coincide with the
spectrum of the corresponding quantum operator $T(f)$ -- ``the range
of values postulate.'' This postulate is violated in our framework.
As we have seen, the classical spin variables are continuous and the
quantum spin operators have discrete spectrum. Nevertheless,
classical averages can be approximated by quantum. Our prequantum
classical statistical model is not about observations, but about
ontic reality (reality as it is when nobody looks at it).

Henry Stapp pointed out [24]: ``The problem, basically, is that to apply
quantum theory, one must divide the fundamentally undefined physical
world into two idealized parts, the observed and observing system,
but {\it the theory gives no adequate description of connection
between these two parts.} The probability function is a function of
degrees of freedom of the microscopic observed system, whereas the
probabilities it defines are probabilities of responses of
macroscopic measuring devices, and these responses are described in
terms of quite different degrees of freedom.'' Since we do know yet from physics  so
much about features of classical $\to$ quantum correspondence map $T,$
we have the freedom to change some conditions which were postulated
in the known no-go theorems -- for example, the range of values
condition. Rejection of this assumption is quite natural, since, as
was pointed by Stapp, a classical variable $f$ and its quantum
counterpart $T(f)$ depend on completely different degrees of
freedom.

\subsection{Is prequantum classical statistical field theory nonlocal?}

As we have seen, PCSFT does not contradict to the known no-go theorems,
in particular, to Bell's theorem. Therefore this theory might be  local. 
However, it is not easy to formulate the problem of
locality/nonlocality in the PCSTF-framework. It is not about
observations. Thus we could not apply Bell's approach  to locality as
locality of observations. On the other hand, on the ontic level
PCSTF operates not with particles, but with fields. At the first
sight, such a theory is nonlocal by its definition, since fields are
not localized. But in field theory there was established a different
viewpoint to locality and we know that both classical and quantum
field theories are local. To formulate the problem of locality for
PCSTF, we should proceed in the same way. Therefore we should
develop a relativistic version of PCSTF. There are some technical
and even ideological problems. As we know, relativistic quantum
mechanics is not a well established theory (at least this is a
rather common opinion). Thus it is meaningless  to develop a
relativistic variant of PCSTF  which would reproduce relativistic
quantum mechanics. The most natural way of development is  to
construct a kind of PCSTF  not for quantum mechanics, but for
quantum field theory and study the problem of locality in such a
framework. It is an interesting and complicated problem which will
be studied in coming papers of the author.

\subsection{Classical statistical model for QFT}

The scheme (\ref{DG}), (\ref{DG1}) is not 
the final scheme of descriptions of physical reality.

It is well known that deformation quantization 
can be performed not only on the finite dimensional, but even on infinite dimensional phase space -- so called 
{\it second quantization,} 
see e.g. [20] (directly in the case of superfields). In this way one finds the correspondence principle
between quantum field theory and classical  field theory, see [20] for the rigorous mathematical
formulation of this principle.  We again use the Planck constant as a small parameter.

On the other hand, in [21] we performed 
dequantization of  quantum field theory. The corresponding classical statistical model
is based on functionals of classical fields, $\psi \to F(\psi).$ Averaging in QFT can be 
considered as an approximation of classical averaging on the space of field functionals.

Combining dequantization for QFT and the correspondence principle for second quantization, we obtain the 
following scheme of descriptions of physical reality:
\begin{equation}
\label{DG1A}
\rm{Theory\; of \; functionals \; of\; classical \;fields}\to  \rm{Quantum\;\;field \; theory}
\end{equation}
\begin{equation}
\label{DG2A}
 \to
\rm{Classical \;field\; theory}
\end{equation}
The crucial point is that the time scales for classical field theory in  two schemes (\ref{DG}),
(\ref{DG1}) and (\ref{DG1A}), (\ref{DG2A})
are different. In the first one we consider the quick time scale:
\begin{equation}
\label{DG1B}
\rm{Classical \;field\; theory}_{\rm{quick \;time}} \to \rm{Quantum\;\;mechanics}
\end{equation}
\begin{equation}
\label{DG3A}
 \to
\rm{Mechanics\; \;of\; classical \; particles}
\end{equation}
In the second one we consider the slow time scale:
\begin{equation}
\label{DG4A}
\rm{Theory\; of \; functionals \; of\; classical \;fields}\to  \rm{Quantum\;\;field \; theory}
\end{equation}
\begin{equation}
\label{DG5A}
\to
\rm{Classical \;field\; theory}_{\rm{slow\; time}}
\end{equation}
In principle we can proceed as long as we like. In [22] I proposed so called {\it third quantization,} 
deformation quantization for functionals  of classical fields. In this way one can construct 
quantum theory of functionals  of classical fields. This theory also can be dequantized and so on.

\medskip

{\bf Conclusion.} {\it We developed consistently the viewpoint on QM as an approximative 
theory\footnote{Cf. with the approach based on the nonlinear Schr\"odinger equation 
[25]--[28]; see [29], [30] for the nonlinear Schr\"odinger equation in the PCSFT-framework.}
for calculation of averages with respect to classical random fields. The parameter of the 
asymptotic expansion for classical averages is determined by two time scales, the quantum time scale
(the atomic time scale in our terminology) and the prequantum time scale 
(the Planck time scale in this paper).\footnote{This paper was strongly motivated by the paper 
of G. F.R. Ellis and T. Buchert [31] on the role of scaling in general relativity and
cosmology. We start with some citations from this paper: "Any
mathematical description of a physical system depends on an {\it
averaging scale} characterizing the nature of the envisaged model.
This averaging scale is usually hidden from view: it is taken to be
understood." "A large scale smoothed-out model of the universe
ignores small scale inhomogeneities, but the averaging effects of
those inhomogeneities may alter both observational and dynamical
relations at the larger scale."  In our approach we discuss
"forbidden scaling", namely scaling beyond the Planck time.
Moreover, we claim that QM can be considered as "a large scale
smoothed-out model of the universe" which ignores small (prequantum)
scale fluctuations.} The first one is a slow time scale and the second one is  
a quick time scale. We presented a classical model, PCSFT, which
gives the possibility to go beyond QM and beyond the Planck time
scale.}

\medskip

I would like to thank B\"orje Nilsson for numerous discussions 
on  scale-analysis in classical and quantum physics and G. `t Hooft
and A. Leggett for discussions on possibilities to go beyond QM.
\medskip

{\bf References}

[1] Khrennikov A. Yu.,   Prequantum classical statistical model with
infinite dimensional phase-space, {\it J. Phys. A: Math. Gen.}, {\bf
38}, 9051-9073 (2005);   Generalizations of Quantum Mechanics
Induced by Classical Statistical Field Theory. {\it Found. Phys.
Letters}, {\bf 18}, 637-650 (2005).

[2] De la Pena  L. and  Cetto A. M., {\it The Quantum Dice: An
Introduction to Stochastic Electrodynamics} ( Kluwer, Dordrecht)
1996.

[3]  Boyer T. H., {\it A Brief Survey of Stochastic
Electrodynamics} in Foundations of Radiation Theory and Quantum
Electrodynamics, edited by Barut A. O. (Plenum, New York) 1980;
Boyer T. H., Timothy H., {\it Scientific American}, pp. 70-78, Aug
1985; see also an extended discussion on vacuum fluctuations in: M.
Scully O., Zubairy M. S., {\it Quantum Optics} (Cambridge University
Press, Cambridge) 1997; Louisell W. H., {\it Quantum Statistical
Properties of Radiation}  (J. Wiley, New York) 1973; Mandel L.  and
Wolf E., {\it Optical Coherence and Quantum Optics} (Cambridge
University Press, Cambridge) 1995.

[4] Cavalleri G., {\it Nuovo Cimento} B, {\bf 112}  (1997) 1193.

[5] Zecca A.  and  Cavalleri G., {\it Nuovo Cimento} B, {\bf 112},
(1997) 1.

[6] Cavalleri G.  and Tonni E., "Discriminating between QM and
SED with spin", in C. Carola and A. Rossi, {\it The Foundations of
Quantum Mechanics (Hystorical Analysis and Open Questions)} (World
Sceintific Publ., Singapore), p.111, 2000.

[7] Nelson  E., Quantum fluctuation, Princeton Univ. Press,
Princeton, 1985.

[8] Davidson M.,  J. Math. Phys. 20  (1979) 1865; Physica A  96 (1979)
465.

[9] Casado A., Fernandez-Rueda A., Marshall T., Risco-Delgado R.,
Santos E., {\it Phys. Rev.} A, {\bf 55} (1997) 3879.

[10] Casado A.,  Marshall T., Santos E., {\it J. Opt. Soc. Am.} B,
{\bf 14} (1997) 494.

[11] Brida G., Genovese M., Gramegna M., Novero C., and Predazzi E.,
{\it Phys. Lett} A, {\bf 299} (2002) 121.

[12] Hiley B., Phase space distribution of quantum phenomena. In {\it
Quantum Theory: Reconsideration of Foundations-2,} Ser. Math.
Modeling, vol. 10, V\"axj\"o University Press, V\"axj\"o, 2003, pp.
267--299.

[13] Khrennikov A. Yu.,  Annalen  der Physik  12  (2003) 575; J.
Phys.A: Math. Gen. 34  (2001) 9965; Il Nuovo Cimento B  117 (2002)
267;  J. Math. Phys. 43  (2002) 789; Ibid  45   (2004) 902; Doklady
Mathematics 71  (2005) 363.

[14] Ballentine L. E., {\it Quantum mechanics} (Englewood Cliffs,
New Jersey) 1989.

[15] Beltrametti E. G., {\it The Logic of Quantum Mechanics}
(Addison-Wesley) 1981.

[16] De Muynck W. M., {\it Foundations of Quantum Mechanics, an
Empiricists Approach} (Kluwer, Dordrecht) 2002.

[17] Accardi L., {\it Urne e Camaleoni: Dialogo sulla realta, le leggi
del caso e la teoria quantistica} (Il Saggiatore, Rome) 1997.

[18] Accardi L.,  Gang Lu Yun,  Volovich  I., {\it Quantum Theory and Its Stochastic Limit}
(Springer, Berlin-Heidelberg) 2006.

[19] Genovese M., {\it Phys. Rep.}, {\bf 413} (2005) 319.

[20] Khrennikov A.Yu., {\it Supernalysis} (Nauka, Fizmatlit, Moscow,
1997, in Russian; English translation: Kluwer, Dordreht, 1999);

[21] A. Yu. Khrennikov, On the problem of hidden variables for quantum field theory, 
{\it Nuovo Cimento} B, {\bf 121}  (2006) 505.

[22] Khrennikov A.Yu., Infinite-dimensional pseudo-differential
operators, {\it Izvestiya Akademii Nauk USSR, ser.Math.},  {\bf 51}
(1987) 46.

[23] Einstein A., {\it The collected papers of Albert Einstein,}
Princeton Univ. Press, Princeton, 1993.

[24] Stapp H. P., $S$-matrix interpretation of quantum theory, {\it Phys. Rev.} D.
{\bf 3}  (1971) 1303.

[25] Bialynicki-Birula I. and Mycielski  J., {\it Annals of Physics}, {\bf 100}
(1976) 62.

[26] Shimony  A., {\it Phys. Rev.}, A {\bf 20}  (1979) 394.

[27] Shull C. G., Atwood D. K., Arthur J., and Horne M. A, {\it Phys. Rev.
Lett.}, {\bf 44} (1980) 765; Weinberg S., {\it Phys. Rev. Lett.}, {\bf 62}  (1989) 485;
Gisin N., {\it Hel. Physica Acta}, {\bf 62} (1989) 363.

[28] Doebner  H. D., {\it Non-linear partial differential operators and
quantization} (Berlin, Springer-Verlag, 1983).

[29] Khrennikov A. Yu., Prequantum classical statistical field theory:
Complex representation, Hamilton-Schr\"odinger equation, and
interpretation of stationary states, {\it Found. Phys. Lett.}, {\bf
19}  (2006) 299.
 
[30] Khrennikov  A. Yu., Nonlinear Schrödinger equations from prequantum
classical statistical field theory, {\it Physics Letters A}, {\bf
357} (2006) 171.

[31] Ellis G. F.R.  and Buchert T., The universe seen at different scales, 
{\it Phys. Lett.}, A {\bf 347}  (2005) 38. 

\end{document}